\documentclass[12pt,preprint]{aastex}

\shorttitle{SN associated with GRB 031203}
\shortauthors{Cobb et al.}

\begin{document}

\title{Discovery of a supernova associated with GRB
031203: SMARTS Optical-Infrared Lightcurves from 0.2 to 92 days}

\author{B.~E. Cobb\altaffilmark{1}, C.~D. Bailyn\altaffilmark{1},
P.~G. van Dokkum\altaffilmark{1}, M.~M. Buxton\altaffilmark{1}, and J.~S. Bloom\altaffilmark{2,3}}
\email{cobb@astro.yale.edu}

\altaffiltext{1}{Department of Astronomy, Yale University, P.O. Box 208101, New Haven, CT 06520.}
\altaffiltext{2}{Harvard-Smithsonian Center for Astrophysics, MC 20, 60 Garden Street, Cambridge, MA 02138}
\altaffiltext{3}{Harvard Society of Fellows, 78 Mount Auburn Street, Cambridge, MA 02138}

\begin{abstract}
Optical and infrared monitoring of the afterglow site of gamma-ray
burst (GRB) 031203 has revealed a brightening source embedded in the
host galaxy, which we attribute to the presence of a supernova (SN)
related to the GRB (``SN 031203''). We present details of the discovery
and evolution of SN 031203 from 0.2 to 92 days after the GRB, derived
from SMARTS consortium photometry in I and J bands. A template type Ic lightcurve,
constructed from SN 1998bw photometry, is consistent with the peak
brightness of SN 031203 although the lightcurves are not identical. 
Differential astrometry reveals that the SN, and hence
the GRB, occurred less than $300 h_{71}^{-1}$ pc (3 $\sigma$) from the
apparent galaxy center.
 The peak of
the supernova is brighter than the optical afterglow suggesting that this source
is intermediate between a strong GRB and a supernova.
\end{abstract}

\keywords{gamma rays: bursts --- supernovae: general --- supernovae: individual (SN 031203)}

\section{Introduction}
Since the discovery of GRB afterglows, the evidence for a physical
connection between gamma-ray bursts (GRBs) and core-collapse
supernovae (SNe) has mounted \citep[see reviews by][]{vanp99,mez01}.
Particularly compelling were observations of
lightcurves and broadband photometry of SN-like features embedded in
GRB afterglow light \citep[see][]{blo03}. Recently, spectroscopic
evidence \citep{Stanek03,Hjorth03,kdw+03} confirmed 
that GRBs are produced in the death of massive stars
\citep{woo93}. To date, SN signatures have been reliably found in only a
few GRBs (see \citealt{blo03}) necessitating the search for and the study of
new GRB-related SNe.

GRB 031203 triggered the IBIS instrument onboard the {\it Integral}
satellite on 3 December 2003 at 22:01:28 UT \citep{Gotz03},
leading to quick discoveries of X-ray \citep{Camp03} and radio
afterglows \citep{Frail03, SKF03}. Spectroscopy of the host galaxy coincident with the radio
transient yielded a redshift of $z=0.1055$ \citep{Proch04},
likely the redshift of the burst itself.
The low redshift (second only to the unusual GRB 980425) of GRB 031203
presents a rare opportunity to create a well-sampled SN lightcurve
using modest aperture telescopes. We began our observations of the field 5 hours
after trigger and continued monitoring periodically for several months.
We reported our discovery of an increase in brightness of the aperture
magnitude of the host, and suggested the emergence of a supernova
was responsible \citep{Bailyn03}. Hereafter, since the
explosion date of the SN is likely that of the GRB, we designate the
SN associated with GRB 031203 as ``SN 031203''. Monitoring of the SN
by other groups has now confirmed the presence of SN 031203 both
photometrically \citep{Bersier04} and spectroscopically \citep{Tag04}.

In this paper we present optical and infrared data obtained with the
SMARTS 1.3m telescope and ANDICAM instrument between 0.2 and 92 days
after the detection of GRB 031203. Observations and data reduction are
reported in section 2. Section 3 describes the aperture photometry and
image subtraction carried out on this data. The resultant evidence of
a SN associated with GRB 031203 is presented. A comparison between
this SN and SN 1998bw is made in section 4. 

\section{Observations and Data Reduction}
Observations commenced at 4 December 2003 3h00m UT, approximately 5 hours
after the {\it Integral} detection of the long-duration (20 sec) GRB 031203
\citep{Gotz03} and follow-up imaging continued for the next 3
months. Data were obtained using the ANDICAM instrument mounted on the 1.3m
telescope at Cerro Tololo Inter-American Observatory (CTIO).\footnote{http://www.astronomy.ohio-state.edu/ANDICAM}
This
telescope is operated as part of the Small and Moderate Aperture Research
Telescope System (SMARTS) consortium.\footnote{http://www.astro.yale.edu/smarts}
The ANDICAM detector consists of a dual-channel camera that allows for
simultaneous optical and IR imaging. The Fairchild 447 2048 $\times$ 2048
optical CCD has a 6\farcm3 $\times$ 6\farcm3 field of view, while the Rockwell 1024 $\times$
1024 HgCdTe ``Hawaii'' IR Array has a 2\farcm4 $\times$ 2\farcm4 field of view. Both
optical and IR images are double-binned in software to give an optical
pixel scale of 0.27 arcsec/pixel and an IR pixel scale of 0.37
arcsec/pixel. While standard optical integrations are underway, the
ANDICAM instrument allows IR images to be ``dithered'' by the slight
adjustment of three tilt axes of an internal mirror.

As GRB 031203 was at low Galactic latitude and thus
subject to high extinction, only I-band data were obtained in the optical.
J-band data were obtained simultaneously. A
combination of 7 telescope re-points and 5 internal dithers were used to
obtain 7 separate 360-second I-band images and 35 separate 60-second
J-band images per data set. Standard reduction was performed on the
I-band images, including overscan bias subtraction, zero subtraction and
flat fielding. The 7 I-band images were then aligned and averaged to
produce a single master I-band frame.

IR flats were taken at two different sky brightnesses.  
Flats with a ``bright'' sky level were median combined
to produce a bright flat frame, and flats with a ``dim''
sky level were median combined to produce a dim flat frame.
The master flat field was produced by subtracting
the dim flat frame from the bright flat frame.
Each J-band image was divided by the 
normalized master flat field. Five sky frames were then produced, one for
each dither position. Each sky frame was formed by median combining
sets of 7 images taken at a given dither position. Median combining produced
star-free sky frames since each of the 7 images at that dither position
were taken at a slightly different telescope position.  Corresponding
sky frames were subtracted from each image with rescaling to
compensate for changes in brightness. Finally, all 35 sky-subtracted
images were aligned and averaged to produce a single master J-band
frame.

\section{Data Analysis}
The discovery of SN 031203 \citep{Bailyn03} was made noting a
differential brightening in the aperture magnitude about the apparent
host from day 0.2 to day 8.3. We have analyzed the full dataset using
both aperture photometry and image-subtraction photometry and find that
both methods give consistent results for the SN lightcurve.

\subsection{Aperture Photometry}
Seeing-matched aperture photometry of the host galaxy of GRB 031203
\citep{Proch04} was performed. The seeing
was matched to a FWHM of $\sim1.2\arcsec$. The relative magnitude of the
host was determined by comparison with 12 on-chip, non-variable,
``standard'' objects. The aperture radius used was 1\farcs9 in I
and 2\farcs6 in J and was chosen to enclose all light from the galaxy
significantly above the sky background level. Relative magnitudes were converted to apparent
magnitudes by comparison, on photometric nights, with the 8 Landolt
standard stars in the field of Rubin 149 (Landolt 1992) for the I-band
images, and with the Persson IR standard stars T832-38078, LHS2397a,
P9106, and P9150 \citep{Per98} for the J-band images.

The resultant lightcurves are shown in Figure 1. 
Both lightcurves clearly reveal the rise and then decay
of a supernova, with a peak between 26 and 34 days. The apparent
magnitude of the host galaxy without the SN component is $19.21\pm0.01$ in I and
$18.29\pm0.03$ in J. At peak, the SN increased the total brightness of the galaxy
by $0.22\pm0.03$ mag in I and $0.29\pm0.04$ mag in J.
Thus the combined light reddens as it approaches maximum
brightness.  While GRB afterglow is described by an achromatic power-law decay,
color-evolution has been associated with the emergence of a SN component
from the afterglow of GRBs, including GRB 030329 \citep{Bloom04}.

The uncertainty in these measurements is determined from the
statistical fluctuation in the measured magnitudes of a non-variable
object with a brightness similar to that of the host galaxy. The
relative error is 0.03 magnitudes in I and 0.06 magnitudes in J.
A number of larger errors (0.09 mag) in J are due to
technical problems that resulted in master J frames produced 
from fewer than 35 individual J images. 
Telescope movement glitches and
a malfunction in the IR array also rendered several IR data sets
unusable and hence only 23 data points are determined in J while 31 data
points are determined in I. An additional uncertainty of 0.07 magnitudes
in I and J exists in the transformation of relative to apparent
magnitudes.

\subsection{Image Subtraction Photometry}
Spatially variable kernel-convolved image subtraction was carried out on the I-band images
using ISIS (Alard 2000). The reference frame that was used 
for subtraction was formed from the 5 I-band images taken more than 
50 days after the GRB, when the SN had faded. The residual
light from the SN near peak brightness is clearly evident in
the subtracted images.  Figure 2 is a ``before and after'' 
example of a typical result of this image subtraction.  

The centroid of the light from the supernova after subtraction
of the galaxy is $0.041\pm0.049$ arcseconds west and
$0.054\pm0.062$ arcseconds south of the center of the host
galaxy. The position of the SN is, therefore, consistent
with the center of the host galaxy.  At the distance of the host galaxy, one arcsecond is
$1.91\,h_{71}^{-1}$ kpc in projection, with ($H_0 = 71 h_{71}$ km
s$^{-1}$ Mpc$^{-1}$). Thus the SN (and by extension the GRB) occurred
within 300 $h_{71}^{-1}$\, pc (3 $\sigma$) of the apparent host
center. Only GRB 970508 occurred closer to the host center
\citep{bkd+02}.

\section{Comparison with SN 1998bw}

As the SN associated with GRB 031203 is reminiscent of SN 1998bw, a comparison
of their lightcurves is in order. SN 1998bw was observed at a lower redshift
and without significant background contribution from its host galaxy. 
The Galactic extinction-corrected lightcurve of SN 1998bw \citep{Galama98}
must, therefore, be shifted from $z=0.0085$ to $z=0.1055$ and must also be added to
the light of the host galaxy of GRB 031203.  

Transforming the lightcurve of SN 1998bw requires a k-correction, stretching, and dimming due both to
the change in luminosity distance and to extinction.  
Wavelengths emitted between the R and I-bands at $z=0.1055$
are redshifted into the observed I band.  The lightcurve of SN 1998bw at a wavelength
between the R and I-bands is simply determined by averaging its R and I-band magnitudes.
This simple k-correction is justifiable as the brightness of SN 1998bw was almost
identical in these two bands, so it can be assumed that the SN's SED was fairly
flat in this wavelength region.

Given the small change in redshift, the wavelength stretching due to redshift
only dims SN 1998bw by 0.1 mag.  The largest dimming 
is due to the change in luminosity distance ($D_L$). 
Assuming a cosmology of $H_o=70$, $\Omega_\Lambda=0.7$
and $\Omega_M=0.3$, then $D_L (z=0.0085) = 36.6$ Mpc and
$D_L (z=0.1055) = 487.4$ Mpc.  The increase
in luminosity distance results in 5.6 magnitudes of dimming.    
SN 1998bw is additionally dimmed by 1.4 mag due to the line of sight 
Galactic extinction toward the host galaxy of GRB
031203.  This value is determined from the Galactic reddening of $E(B-V)=0.78$ toward the
host galaxy \citep{Proch04} and assuming the Galactic extinction curves of \citet{CCM89}.
This calculation assumes that both SN have undergone a similar amount
of host galaxy extinction, and, therefore, no attempt is made to
correct for such extinction.
The last step is to add the light of the host galaxy, assuming the
host galaxy has an I magnitude of 19.21.

Despite the uncertainties inherent to the above procedure,
once SN 1998bw is shifted into the host galaxy of GRB 031203 only
an additional 0.02 magnitudes has to be added to its lightcurve
for its peak to match that of SN 031203.    
Figure 3 shows this lightcurve as a solid line overlaid on the lightcurve
of SN 031203. Clearly SN 1998bw peaks earlier than SN 031203.  However,
stretching the lightcurve of SN 1998bw makes the decline much too long compared
to our data.  The dotted line in Figure 3, for example, shows a stretch of 1.7, chosen
to make the highest points coincide. The problems with the decline are clear; 
indeed, no combination of stretch and offset of the SN lightcurve fits the data.
This difficulty was not apparent in the data of \citet{Thomsen04} who had somewhat
sparser sampling.

Given the inconsistencies of the template Ic and the ANDICAM data, we
cannot exclude the possibility that the brightening source was due to
another type of supernova (although \citet{Tag04} report spectroscopic
evidence of a Ic origin). However, local Ic SNe show a
variety of rise and fall timescales as well as a large range in
brightness distributions \citep[e.g.][]{mdm+02} so the differences
may simply be inherent in the GRB-related SNe. To be sure, one of the
more puzzling emergent trends in GRB-related SNe is why the peak
brightnesses of the SNe should be so similar ($M_V \approx -19.5$ mag)
\citep{ZKH04,Bloom04}, yet the lightcurves differ substantially. Since the peak
brightness scales roughly as the mass of the synthesized $^{56}$Ni
\citep[whereas the timescale depends on the mass of the ejecta and the
explosion energy ---][]{Nomoto03}, this
trend may point to a regularizing mechanism for $^{56}$Ni synthesis in
SN-GRBs perhaps related to the apparent regularization of energy
release in the prompt burst phase \citep{fks+01}.

We note that there is scant evidence for the existence of the afterglow
itself.  The peak luminosity of the SN is, therefore, greater than that
of the afterglow just a few hours after the GRB.  This is in 
stark contrast to GRB 030329 in which the afterglow after 5 hours
was over 5 magnitudes brighter than the peak of the supernova \citep{Lipkin03}.
It is interesting to note that the fluence of the GRB itself (corrected
for luminosity distance) was over 3 orders of magnitude larger in the case
of GRB 030329 than GRB 031203 \citep{Ricker03,Hurley03}.  This may suggest that there is a large
population or perhaps a continuum of events intermediate between
GRBs and SNe.  If so, Swift may reveal a population of faint GRBs
whose late-time optical lightcurves are dominated by a supernova,
as is the case reported here.

\acknowledgments
This work was partially supported by NSF grant AST 0098421 to C.~D.~B..
J.~S.~B.~is supported by the Harvard Society of Fellows and by a
generous research grant from the Harvard-Smithsonian Center for
Astrophysics.  We thank David Gonzalez and Juan Espinoza for their dedication
to observing this source, and Suzanne Tourtellotte for assistance
in the reduction of the optical data.

\clearpage

\begin{figure}
\includegraphics[width=1\textwidth]{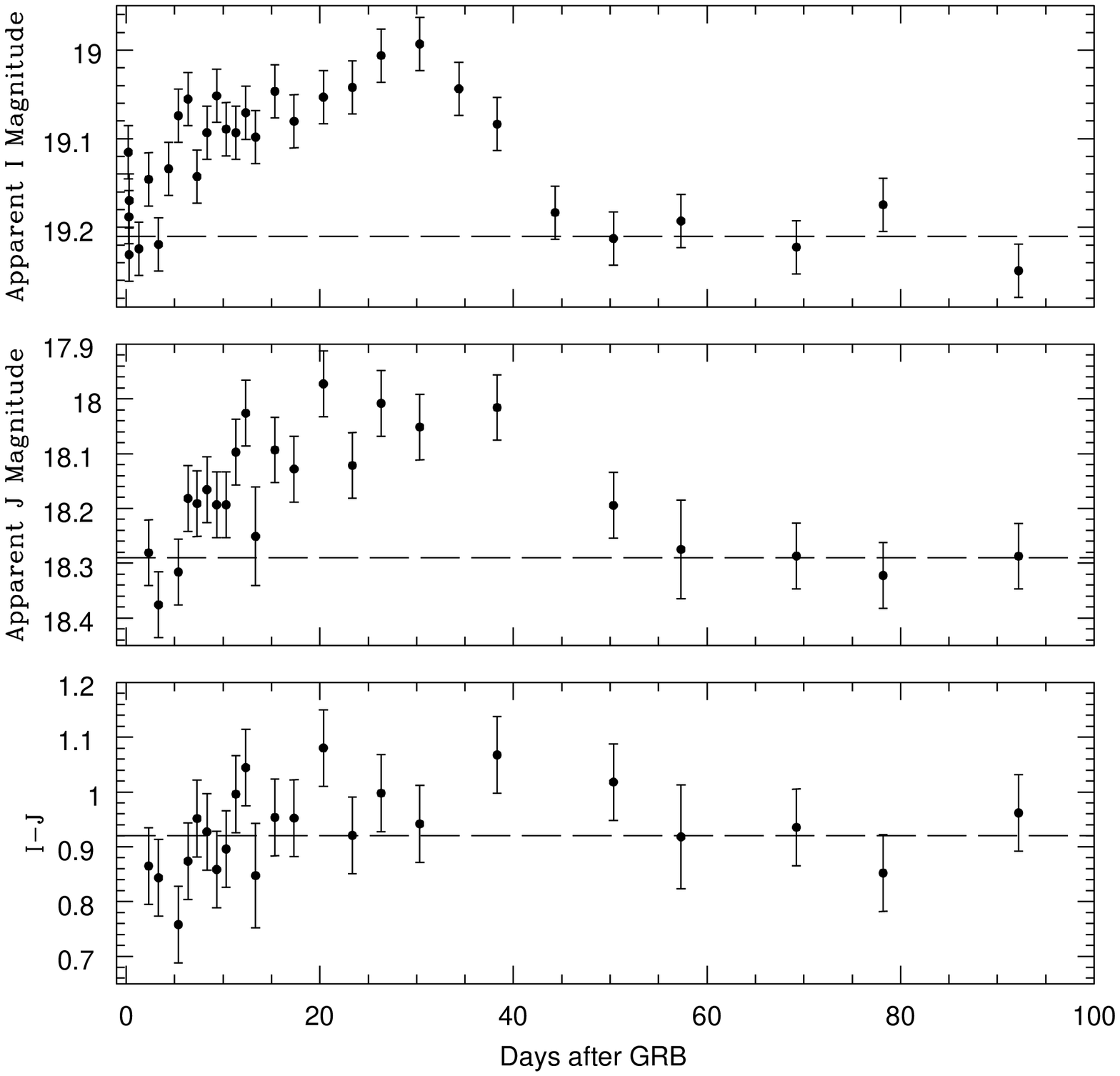}
\caption{\textit{Top panel:} The aperture photometry I-band lightcurve of the host+SN+afterglow of GRB 031203. The
dashed line shows the I magnitude of the host (19.21).
\textit{Middle panel:} The aperture photometry J-band lightcurve of the host+SN+afterglow of GRB 031203. 
The dashed line shows the J magnitude of the host (18.29).
The rise and decay in brightness is consistent with a type Ic SN. 
These values have not been corrected for Galactic extinction. 
\textit{Bottom panel:} I-J color evolution, a slight reddening in color is noted as
the SN approaches peak brightness.  The dashed line shows the I-J color of the host (0.92).}
\end{figure}

\begin{figure}
\includegraphics[width=1\textwidth]{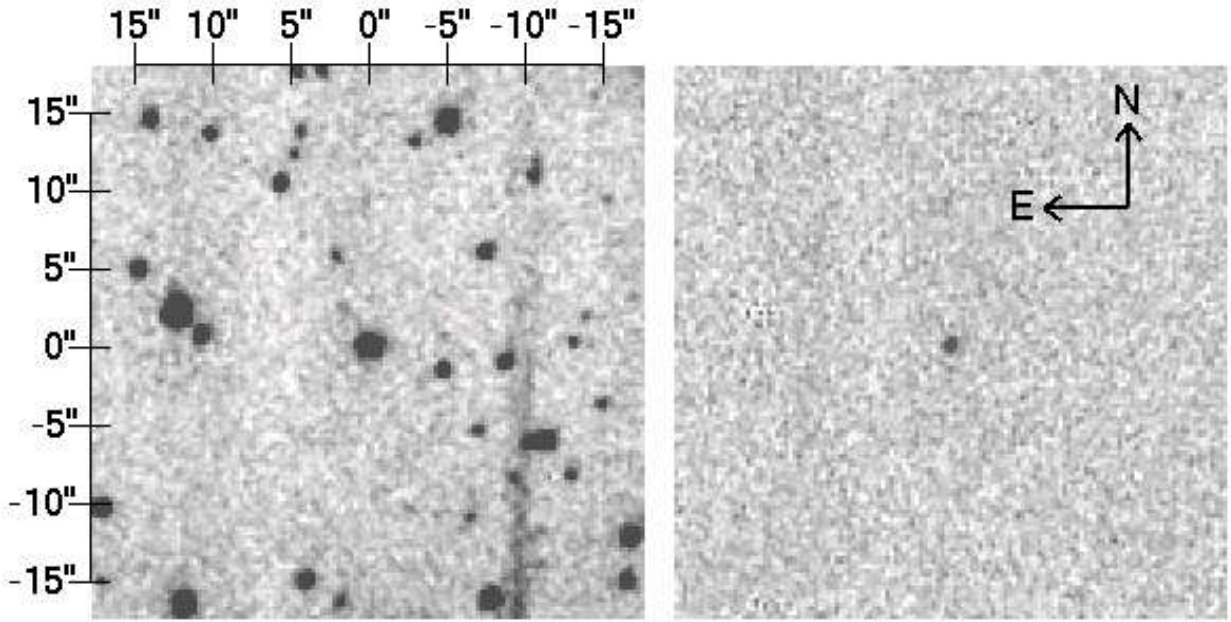}
\caption{Day 26 image before and after kernel-convolved image subtraction with ISIS. The object in the
center of the unsubtracted $35\arcsec \times 35\arcsec$  field is the host galaxy of GRB 031203, which
is extended in the EW direction. When a reference frame 
is subtracted from the image, the residual light from the SN is clearly evident. The reference
frame was formed from the 5 I-band images taken more than 50 days after the GRB, when the SN had
faded.} 
\end{figure}

\begin{figure}
\includegraphics[width=1\textwidth]{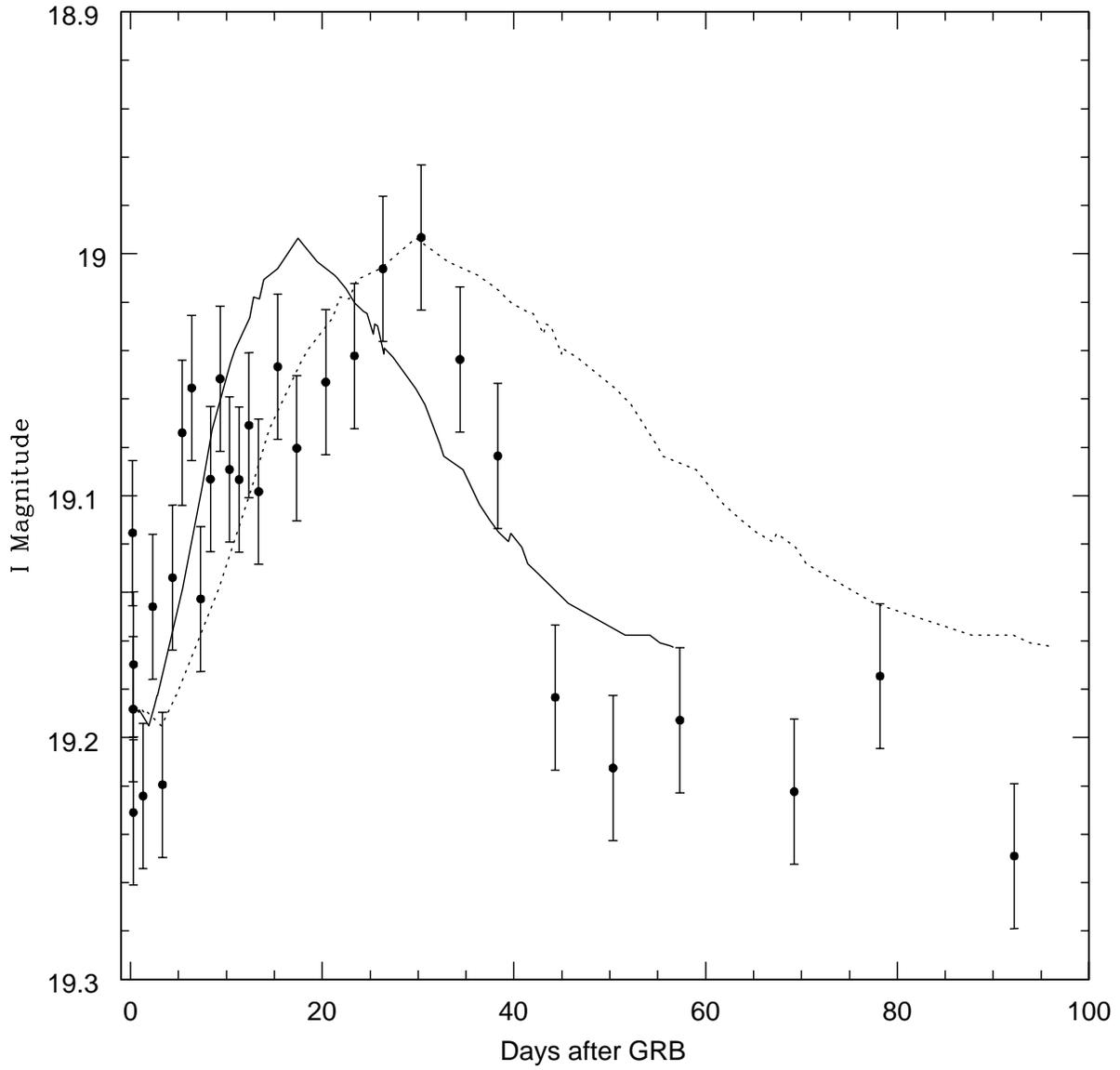}
\caption{The lightcurve of SN 1998bw moved to $z=0.1055$ and placed in the host galaxy of GRB 031203 
(solid line) is shown overplotted on the lightcurve
of the host galaxy (points). An additional 0.02 mag has been added to the shifted 
SN 1998bw lightcurve so that the SNe reach the same peak brightness. If a stretch
of 1.7 is applied (dotted line) the peaks coincide but the declines are
inconsistent.}
\end{figure}

\end{document}